\documentclass[aps,twocolumn,floatfix,superscriptaddress,amssymb]{revtex4}

\usepackage{braket}
\usepackage{amsmath}
\usepackage{xfrac}
\usepackage[normalem]{ulem}
\usepackage{xcolor,color}
\usepackage{graphicx}% Include figure files
\usepackage{dcolumn}% Align table columns on decimal point
\usepackage{bm}% bold math
\begin{document}

\preprint{APS/123-QED}

\title{Third-Order Exceptional Point in Non-Hermitian Spin-Orbit-Coupled cold atoms}% Force line breaks with \\

\author{Yu-Jun Liu}%
\affiliation{%
 Department of Physics, The Hong Kong University of Science and Technology, Clear Water Bay, Kowloon, Hong Kong, China
}%

\author{Ka Kwan Pak}%
\affiliation{%
 Department of Physics, The Hong Kong University of Science and Technology, Clear Water Bay, Kowloon, Hong Kong, China
}%

\author{Peng Ren}%
\affiliation{Department of Physics and Astronomy, Rice University, Houston, TX, USA}

\author{Mengbo Guo}%
\affiliation{%
 Department of Physics, The Hong Kong University of Science and Technology, Clear Water Bay, Kowloon, Hong Kong, China
}%
% \author{Yu-Jun Liu}%
% \affiliation{%
%  Department of Physics, The Hong Kong University of Science and Technology, Clear Water Bay, Kowloon, Hong Kong, China
% }%

\author{Entong Zhao}%
\affiliation{%
 Department of Physics, The Hong Kong University of Science and Technology, Clear Water Bay, Kowloon, Hong Kong, China
}%

\author{Chengdong He}%
\affiliation{%
 Department of Physics, The Hong Kong University of Science and Technology, Clear Water Bay, Kowloon, Hong Kong, China
}
\affiliation{Department of Physics and Astronomy, Rice University, Houston, TX, USA}

\author{Gyu-Boong Jo}%
\email{gbjo@rice.edu}
\affiliation{%
 Department of Physics, The Hong Kong University of Science and Technology, Clear Water Bay, Kowloon, Hong Kong, China
}
\affiliation{Department of Physics and Astronomy, Rice University, Houston, TX, USA}

\date{\today}

\begin{abstract}
Exceptional points (EPs) has seen substantial advances in both experiment and theory. However, in quantum systems, higher-order exceptional points remain of great interest and possess numerous intriguing properties yet to be fully explored. Here, we describe a \emph{PT} symmetry-protected three-level non-Hermitian system with the dissipative spin-orbit-coupled (SOC) fermions in which a third-order exceptional point (EP3) emerges when both the eigenvalues and eigenstates of the system collapse into one. The band structure and its spin dynamics are explored for $^{173}$Yb fermions. We highlight the enhanced sensitivity to the external perturbation of EP3 with cubic-root energy dispersion. Additionally, we investigate the second-order exceptional point (EP2) with square-root energy dispersion in a three-level quantum system with the absence of parity symmetry, which proves that the enhanced sensitivity closely relates to the symmetries of the NH system. Furthermore, we analyze the encircling behavior of EP3 in terms of the adiabatic limit and the nonadiabatic dynamics and discover some different results from that of EP2.
% Previous version: We describe a three-level non-Hermitian system with the dissipative spin-orbit-coupled (SOC) fermions in which a third-order exceptional point (EP3) occurs when both the eigenvalues and eigenstates of the system collapse into one. The band structure and spin dynamics are explored. We highlight the enhanced sensitivity to the external perturbation of EP3 in comparison to the second-order exceptional point (EP2). Furthermore,  we analyze the differences between EP3 and EP2 in terms of the encircling within the adiabatic limit and the nonadiabatic dynamics of encirclement.
\end{abstract}

\maketitle

%{\color{red}
%\begin{verbatim}
%For citation, please put the space as ~\cite{}.
%\end{verbatim}
%}
Non-Hermitian (NH) quantum systems reveal unprecedented  phenomena near singular points called exceptional points (EPs). At these EPs, both the eigenvalues and eigenstates of the system merge into one within the parametric space~\cite{dembowski2004encircling,doppler2016dynamically,mailybaev2005geometric,ozturk2021observation}. Unlike diabolic points (DPs) which show only eigenvalue degeneracy, EPs feature complete degeneracy where both eigenvalues and eigenstates collapse into a single state. At EPs, intersecting eigensurfaces form Riemann sheets~\cite{doppler2016dynamically,dembowski2001experimental} that exhibit chiral behavior~\cite{dembowski2004encircling}. Especially, second-order exceptional points (EP2) have been demonstrated in two-mode classical systems, including photonics~\cite{miri2019exceptional}, acoustics~\cite{ding2016emergence}, membrane~\cite{xu2016topological}, as well as in two-state quantum systems such as NV centers~\cite{wu2019observation}, trapped ions, superconducting qubits~\cite{naghiloo2019quantum}, and neutral atoms~\cite{ren2022chiral,Zhao.2023,Li.2019}.

%Hamiltonians have been realized and observed in experimental setups involving gain and loss~\cite{lauber1994geometric}, e.g., photonic systems~\cite{miri2019exceptional} and ultracold atom systems~\cite{ren2022chiral,Zhao.2023,Li.2019}. Especially, adjustable parameters of NH system enable us to explore the abrupt phase transition around the vicinity of the singular points, commonly referred to as exceptional points (EPs) where both the eigenvalues and eigenstates of the system collapse into one within the parametric space~\cite{dembowski2004encircling,doppler2016dynamically,mailybaev2005geometric,ozturk2021observation}.

%In contrast to diabolic points (DPs), which are characterized solely by eigenvalue degeneracy, both the eigenvalues and the eigenstates degenerate into one for EPs, wherein the eigenstates collapse into a single state. The formation of intersecting eigensurfaces at EPs involves the Riemann sheets~\cite{doppler2016dynamically,dembowski2001experimental}. Therefore, the encircling behaviors associated with DPs and EPs are fundamentally different. Encircling a DP results in the accumulation of a Berry phase solely by the eigenstates~\cite{berry1984quantal}. On the other hand, a dynamical encirclement in the vicinity of an EP leads to chiral state conversion, which is not observed in the case of DPs~\cite{dembowski2004encircling}. 

Beyond EP2, higher-order exceptional points emerge when there is increased degeneracy of eigenvalues and eigenstates in a multi-state NH system~\cite{mandal2021symmetry}. These higher-order EPs have been extensively studied theoretically~\cite{zhong2018winding, kullig2023higher} and experimentally demonstrated primarily in classical systems, such as photonics~\cite{hodaei2017enhanced,patil2022measuring}, acoustics~\cite{ding2016emergence, tang2020exceptional}, and cavity optomechanics~\cite{aspelmeyer2014cavity, patil2022measuring, guria2024resolving}. Recent developments have extended studies of higher-order EPs to quantum systems, including nitrogen-vacancy centers~\cite{wu2024third}, ion-cavity systems~\cite{kim2023third} and ultracold Bose gas~\cite{pan2019high, pan2019interacting}.

Higher-order EPs exhibit higher sensitivity to small external perturbations, denoted as $\epsilon$ ($\epsilon\ll1$), in their vicinity~\cite{hodaei2017enhanced,wu2021high}. The response is amplified to a magnitude of $\epsilon^{1/N}$, where $N$ denotes the order of the EP~\cite{zhong2018power,lin2016enhanced}. In quantum systems, however, the significance of quantum noise must be considered when evaluating enhanced sensitivity near higher-order EPs in EP-based sensing~\cite{lau2018fundamental,clerk2010introduction}. Beyond sensing applications, higher-order EPs demonstrate distinct encircling behavior compared to EP2~\cite{mailybaev2005geometric,hassan2017dynamically,hassan2017chiral}, exhibiting chiral quantum state conversion~\cite{ren2022chiral,doppler2016dynamically, pick2019robust, liu2021dynamically}.

Here, we propose an experimental scheme using neutral atoms to create a three-level dissipative SOC system with two $\Lambda$-type configurations, featuring adjustable parameters such as two-photon detuning and dissipations for $^{173}$Yb atoms~\cite{ren2022chiral,zhang2018collective,zhang2019recent}. By studying the band structure, we reveal parity-time (\emph{PT}) symmetry breaking across the EP3 in the parameter space. We examine two cases in the three-level systems corresponding to EP2 and EP3, introducing small external perturbations to analyze their effects. Through examining the energy surfaces' splitting scale relative to the applied external perturbation scale, we confirm response amplification as $\epsilon^{1/N}$ in the presence of higher-order EPs. Finally, we numerically analyze the encircling dynamics of EP3 and compare it to the analytic expressions of EP2's encircling process, revealing distinct characteristics in both adiabatic limit and nonadiabatic dynamics specific to EP3.

%Therefore, we aim to extend the theoretical calculations pertaining to the encirclement of EP2 to EP3 scenario~\cite{laha2020third}. As a result, we discover a lower adiabatic limit for encirclement around EP3 in comparison to EP2. Additionally, we observe input-dependent nonadiabatic dynamics for EP3, whereas EP2 is not sensitive to it. {\color{red} However, EP3 has not been realized in a topological band in the quantum regime.}

% Definition of EP2

\vspace{20pt}
\paragraph*{\bf Experimental Scheme.}

\begin{figure}%[htbp]
    \centering
    \includegraphics[width=1\linewidth]{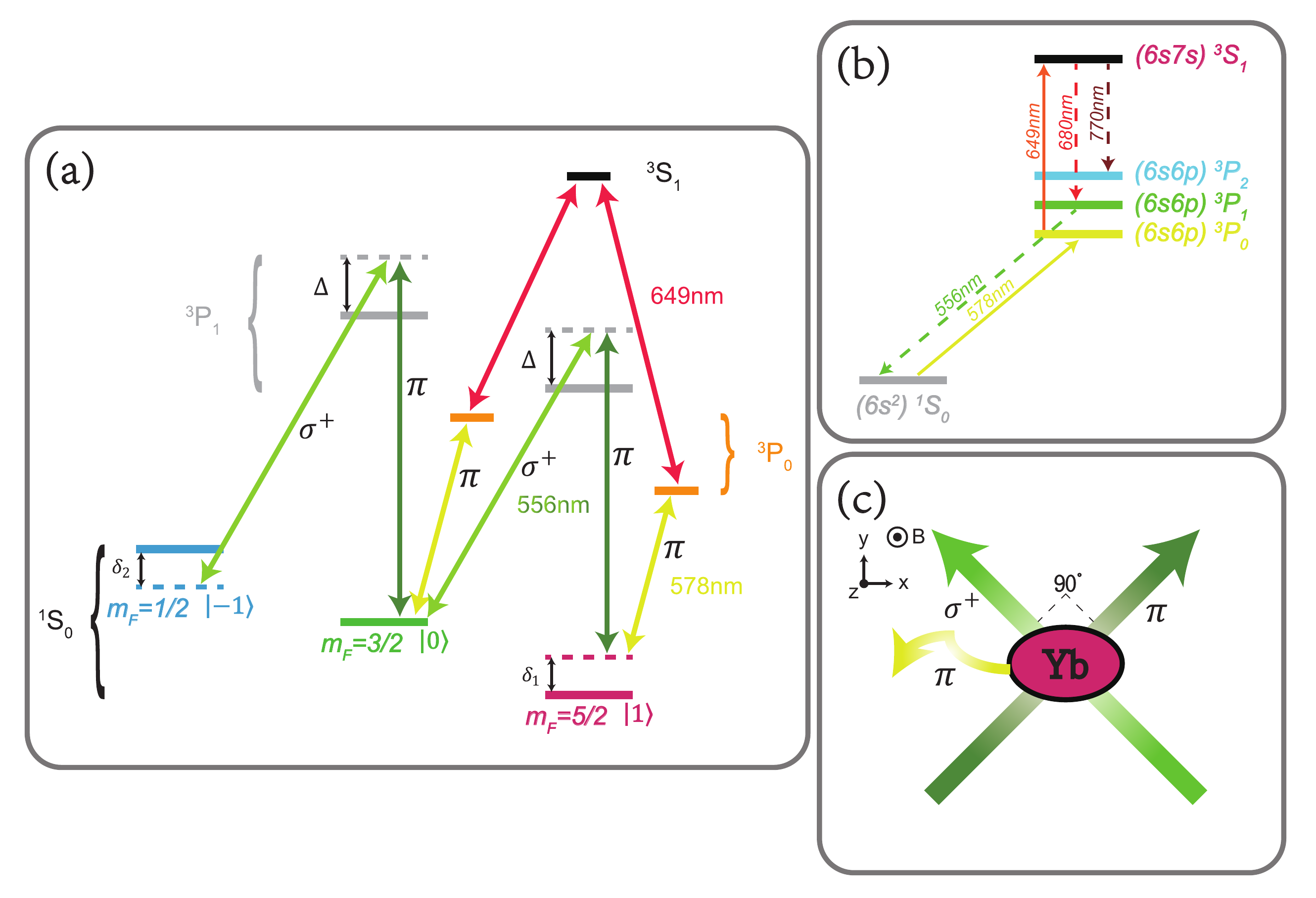}
    \caption{\textbf{Experimental scheme for realizing three-level non-Hermitian spin-orbit-coupled system} (a) shows the relevant energy levels and transitions. Two $556nm$ far-blue-detuned beams are applied to couple three nuclear spins in the successive $\Lambda$-type configurations to form a spin-1 system. The atom loss is induced by the $578$~nm beam corresponding to the clock transition. (b) shows the scheme to kill the remaining $^3P_0$ atoms with long lifetime. Due to the spontaneous decay from $^3S_1$ to $^3P_x$, three laser beams of wavelength $649nm$, $680nm$ and $770nm$ are utilized to ensure the atoms removed from the trap. (c) shows the experimental setup of two beams of perpendicular Raman lights, a $578nm$ loss beam and an electro-optic modulator for generation of two successive $\Lambda$-type configurations.}
    \label{schemefig}
\end{figure}

%For a two-band models, we need to tune three parameters to achieve the DP in Hermitian system. However, the EP2 occurs in two-dimension NH system due to the different formation of the nodal band structure from the Hermitian counterpart. Besides, the nodal NH phase appears even in one dimension for the NH system protected by \emph{PT} symmetry~\cite{}. 

In general, the n$^{th}$-order EP emerges in $(2n-2)$-dimension NH system. Therefore, we need to tune 4 parameters to obtain EP3 in a three-band NH systems. However, the co-dimension of EPs will reduce to two for the NH systems protected by specific symmetries~\cite{budich2019symmetry,mandal2021symmetry}.

To create a EP3, we consider three nuclear spin states in the ground state $^1S_0$ of the ultrocold $^{173}$Yb fermions, which are optically coupled by two successive $\Lambda$-type Raman configurations as described in Fig.\ref{schemefig}(a-c)~\cite{dalibard2011colloquium}. For the sake of convenience, we refer to these three nuclear spin states as $\ket{1}$, $\ket{0}$ and $\ket{-1}$ hereafter. A blue-detuned lift beam with $\sigma^-$ polarization should be applied to separate the other nuclear spins of no interest since there are six nuclear spins in the ground state $^1S_0$ of $^{173}$Yb~\cite{lin2011spin,song2016spin}. 
We can adjust the ac stark shift by appropriately detuning the lift beams to isolate out a three-level subspace composed of $\ket{m_F=\frac{5}{2},\frac{3}{2},\frac{1}{2}}$. Then we establish two successive tunable $\Lambda$-configurations, one involving $\ket{1}$ and $\ket{0}$, and the other involving $\ket{0}$ and $\ket{-1}$~\cite{zhang2018collective}. 

In order to realize the dissipative non-Hermitian system, near-resonance blasting pulses are applied to induce the dissipations in the $\ket{1}$ and $\ket{0}$ states, corresponding to the $^1S_0-^3P_0$ clock transition~\cite{cho2012optical,He.2019}. However, it is essential to remove the atoms remaining in the $^3P_0$ state with a long lifetime extending up to tens of seconds. This can be done by the $^3P_1-^3S_1$ transition, but atoms in the $^3S_1$ state spontaneously decay to the three $^3P_{0, 1, 2}$ states, with a branching ratio $\lambda_0:\lambda_1:\lambda_2=1:3:5$, as illustrated in Fig.\ref{schemefig}(b). By using repump beams at $680$
~nm and $770$~nm (see Fig.~\ref{schemefig}(b)), we can repump atoms back to $^3P_{0}$ ensuring minimal spontaneous decay to the $^1 S_0$ state~\cite{cho2012optical}. 

%\warn{We need to discuss to what extend our system can be effectively described by effective non-Hermitian Hamiltonian, comparecd to Lindblad description.}
In cold atom experiments with enforced post-selection measurements, we observe the averaged quantum trajectory of multiple stochastic processes driven by an effective non-Hermitian Hamiltonian~\cite{ren2022chiral,PhysRevA.108.013302}. Then, we can neglect the jumping term $a_n\rho{a_n^\dag}$ from the Lindblad master equation as follow,
\begin{equation}
    \Dot{\rho}(t)=-i\left(H_{eff}(t)\rho(t)-\rho(t)H_{eff}^\dag(t)\right)+\sum_na_n\rho(t)a_n^\dag
    \label{eq-lindblad}
\end{equation}
where $a_n$ is the jump operator, $H_{eff}=H_0-ia_n^\dag{a_n}$ is the non-Hermitian effective Hamiltonian and $\rho$ is the density matrix of the state of the single particle.
Consequently, we establish a three-level {\it effective} non-Hermitian Hamiltonian~\cite{minganti2019quantum},
\begin{equation}
    \label{hamiltonianeq}
    H_{eff}=
    \begin{pmatrix}
        \frac{\hbar^2(k-2q_r)^2}{2m}-\delta_1&\frac{\Omega_{R1}}{2}&0\\
        \frac{\Omega_{R1}}{2}&\frac{\hbar^2k^2}{2m}&\frac{\Omega_{R2}}{2}\\
        0&\frac{\Omega_{R2}}{2}&\frac{\hbar^2(k+2q_r)^2}{2m}+\delta_2\\
    \end{pmatrix} + H_{loss},
\end{equation}
where $k$ denotes the quasi-momentum of the atoms along $\hat{x}$ direction, $q_r$ represents the recoil momentum given by $q_r=\left(\frac{2\pi}{\lambda_{556nm}}\right)sin(\frac{\pi}{2})$, $m$ refers to the mass of a single $^{173}$Yb atom and $\Omega_{R1}$ ($\Omega_{R2}$) signifies the coupling strength associated $\ket{1}\leftrightarrow\ket{0}$ ($\ket{0}\leftrightarrow\ket{-1}$). It is reasonable to approximate $\Omega_{R1}=\Omega_{R2}$, given a single-photon detuning of approximately 1~GHz~\cite{zhang2018collective}. Furthermore, the individual detunings $\delta_1$ and $\delta_2$ can be independently controlled by introducing an additional $\sigma^-$ polarized lift beam. This beam splits the degeneracy of the hyperfine levels within the ground state $^1S_0$ of $^{173}$Yb. The dissipative process, represented by the term $H_{loss}=-\frac{i}{2}\left(\gamma_1\ket{1}\bra{1}+\gamma_0\ket{0}\bra{0}\right)$, are induced by the nearly resonant blasting pulse associated with the $^1S_0-^3P_0$ clock transition~\cite{cho2012optical,minganti2019quantum}.

\vspace{10pt}
\paragraph*{\bf Exceptional points in dressed bands } 

\begin{figure*}[htbp]
    \centering
    \includegraphics[width=1\linewidth]{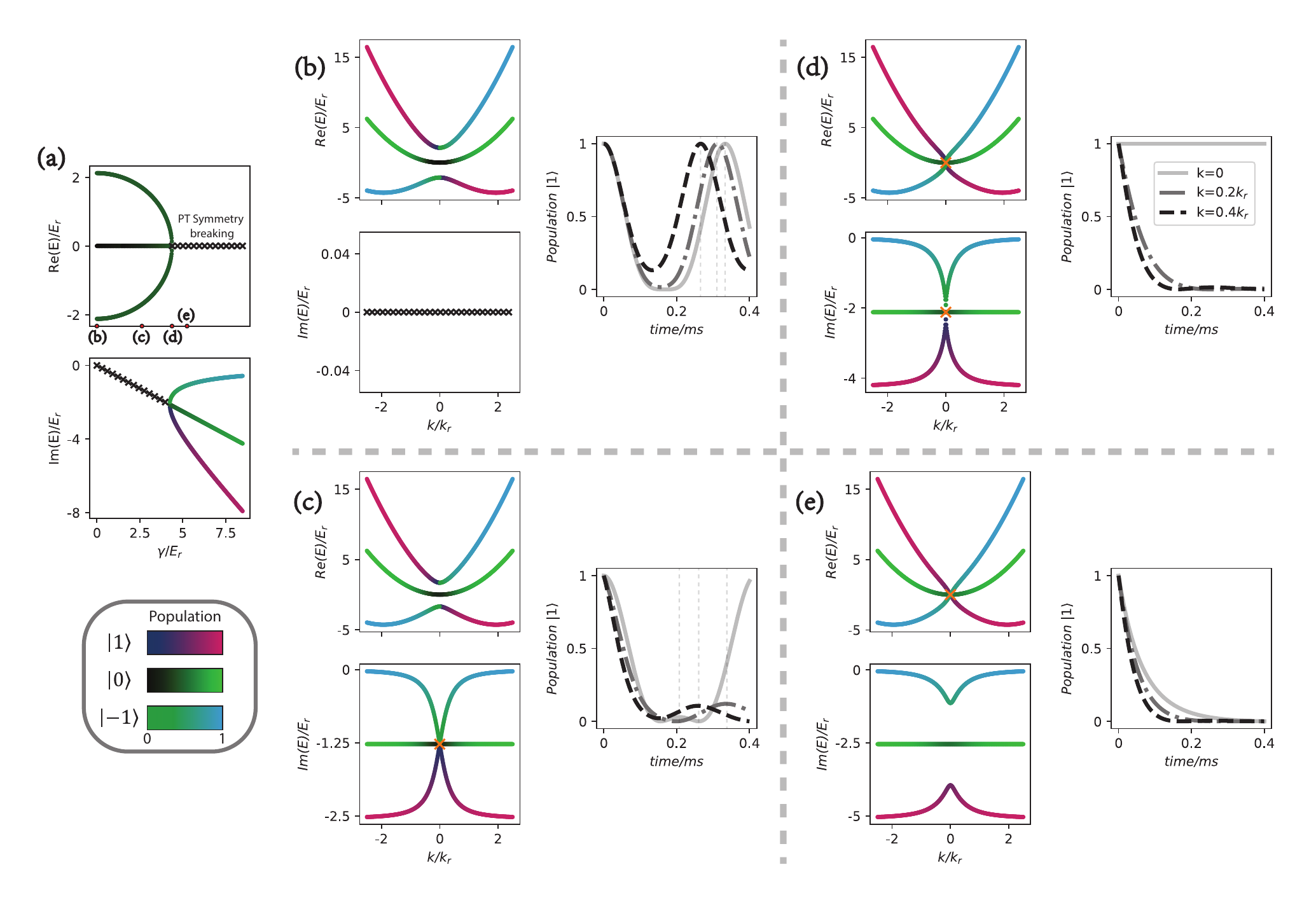}
    \caption{\textbf{Band structure of non-Hermitian spin-1 system} (a) exhibits the phase diagram of the \emph{PT} symmetry breaking transition near EP3 at $k=0$ for $\delta_1=-\delta_2=4E_r$ and $\Omega_{R1}=\Omega_{R2}=3E_r$. (b-e) show the single-particle energy dispersion of dressed states in quasi-momentum space with increasing dissipation at $\gamma_0=0, 0.6\sqrt{2}\Omega_{R1}, \sqrt{2}\Omega_{R1}$ and $1.2\sqrt{2}\Omega_{R1}$. The figures on the right depict the temporal evolution of $\ket{1}$ population at $k=0, 0.2k_r, 0.4k_r$. }
    \label{bandfig}
\end{figure*}

The three-level non-Hermitian Hamiltonian Eq.(\ref{hamiltonianeq}) with the eigenbases ${\ket{1}, \ket{0}, \ket{-1}}$ yields complex eigenvalues $\lambda_{0, \pm1}$ associated with the dressed state $\ket{0', \pm1'}(\Vec{k})$. In the momentum space, the Hamiltonian (\ref{hamiltonianeq}) gives rise to three distinct energy bands referred to as the \emph{upper}, \emph{middle} and \emph{lower} bands, respectively. Fig.\ref{bandfig} illustrates the band structures of the non-Hermitian system in momentum space under different dissipation rates for states $\ket{0}$ and $\ket{1}$.

In the subsequent analysis, we set the recoil energy $E_r=\frac{\hbar^2q_R^2}{2m}$ as the unit for parameters and energy values. For the sake of simplicity, we set $\delta_1=-\delta_2=4E_r$, $\Omega_R=\Omega_{R1}=\Omega_{R2}$, and the Hamiltonian (\ref{hamiltonianeq}) at $k=0$ simplifies to the form:
\begin{equation}
    \label{hamiltoniank0eq}
    H=
    \begin{pmatrix}
        -\sfrac{i\gamma_1}{2}&\sfrac{\Omega_{R}}{2}&0\\
        \sfrac{\Omega_{R}}{2}&-\sfrac{i\gamma_0}{2}&\sfrac{\Omega_{R}}{2}\\
        0&\sfrac{\Omega_{R}}{2}&0\\
    \end{pmatrix}.
\end{equation}

At $k=0$, there exists an EP3 characterized by the conditions $\gamma_1=2\gamma_0$ and $\gamma_0=\sqrt{2}\Omega_{R}$. In this scenario, the eigenvalues and eigenstates of the three energy bands collapse at a single point, denoted as $-\sfrac{i\gamma_0}{2}$ and $(\sfrac{1}{2}, \sfrac{i\sqrt{2}}{2}, -\sfrac{1}{2})^T$, respectively. To observe the closing of the band gap as the system approaches the EP3, we maintain a fixed relation of $\gamma_1=2\gamma_0$ and gradually increase the dissipation rate $\gamma_0$ from $0$ to $1.2\sqrt{2}\Omega_R$, as depicted in Fig.\ref{bandfig}(b-e).

Before delving into the analysis of the energy bands, let us outline the method to visualize the band gap in the real system by extracting the temporal evolution of each spin population~\cite{cohen1986quantum,li2012sum}. Since we prepare a polarized Fermi sea denoted by $\ket{\phi}$ as the initial state of the system, we can expand $\ket{\phi}$ within the eigenbasis of Eq.(\ref{hamiltonianeq}) for a specific quasimomentum $k$ by expressing it in terms of the respective eigenvectors as follow, 
\begin{equation}
    \label{speq}
    \ket{\phi} = \sum_{n'=0, \pm1}C_{n'}\ket{n'}
\end{equation}
where $\sum_{n'=0,\pm1}\left|C_{n'}\right|^2=1$ and $\ket{n'}_{n'=0, \pm1}$ are the eigenvectors dressed by the spin-orbit-coupling corresponding to the relative quasi-momentum $k$. Then the dressed eigenstates can be expressed by the bare spin states as,
\begin{equation}
    \label{eigendresseq}
    \begin{pmatrix}
        \ket{1'}\\
        \ket{0'}\\
        \ket{-1'}\\
    \end{pmatrix}=
    \begin{pmatrix}
        b_{1,1}&b_{1,0}&b_{1,-1}\\
        b_{0,1}&b_{0,0}&b_{1,-1}\\
        b_{-1,1}&b_{-1,0}&b_{-1,-1}\\
    \end{pmatrix}
    \begin{pmatrix}
        \ket{1}\\
        \ket{0}\\
        \ket{-1}\\
    \end{pmatrix}.
\end{equation}

To determine the energy band gap within the momentum-resolved interval, we observe the Rabi oscillation of each spin population at specific quasimomentum by quenching the system. The Rabi oscillation of each spin should follow the relation,
\begin{equation}
    \label{compositioneq}
    \ket{\phi(t)}=\sum_{n'=0,\pm1}C_{n'}e^{-i\omega_{n'}t}\ket{n'},
\end{equation}
where $\hbar\omega_{n'}$ reflects the eigenenergy of each dressed band in Eq.(\ref{hamiltonianeq}). The time evolution of each spin population is thus given by $\left|\braket{n|\phi(t)}\right|^2=\left|\sum_{n'}C_{n'}b_{n',n}e^{-i\omega_{n'}t}\right|^2$~\cite{cohen1986quantum}.

%Since we prepare a Fermi sea denoted by $\ket{\phi}$ as the initial state of the system, we can expand this state for a specific quasimomentum $k$ by expressing it in terms of the respective eigenvectors as follows,
%\begin{equation}
%    \label{speq}
%    \ket{\phi} = \sum_{n=0, \pm1}C_n\ket{r_n, g_n, b_n},
%\end{equation}
%where $\sum_{n=0,\pm1}\left|C_n\right|^2=1$ and $\ket{r_n, g_n, b_n}_{n=0, \pm1}$ are the eigenvectors corresponding to the relative quasimomentum $k$. The parameters within the eigenvectors represent the populations of each spin state: $\ket{1}$, $\ket{0}$ and $\ket{-1}$, respectively.

%To determine the band gaps in momentum space, we can obtain three Rabi frequencies by analyzing the time evolution of each spin population using the quench method within the eigenbasis $\{\ket{0}, \ket{\pm1}\}$. This can be achieved as follows,
%\begin{equation}
%    \label{compositioneq}
%    \ket{\phi(t)}=\sum_{m}\left[\sum_{n=0,\pm1}C_nm_ne^{-i\omega_nt}\right]\ket{m},
%\end{equation}
%where $m={r:1, g:0, b:-1}$ \warn{What does this mean? ; Defind $m_n$}. The time evolution of the population of each spin is given by $\left|\braket{m|\phi(t)}\right|^2=\left|\sum_nC_nm_ne^{-i\omega_nt}\right|^2$, which might compose three Rabi frequencies~\cite{cohen1986quantum}.

With the prior knowledge, we can extract the temporal evolution of $\ket{1}$ population from Eq.(\ref{compositioneq}) at different quasimomenta, namely, $k=0, 0.2k_r, 0.4k_r$.

The dash lines in the evolution graphs of Fig.\ref{bandfig}(b-c) illustrate that the increase in the dissipation rate leads to the smaller energy gaps in real part. It is evident that the introduction of dissipation with a non-zero value of $\gamma_0$ leads to the emergence of spin population damping, which indicates the separated energy band gap in the imaginary part~\cite{ren2022chiral}.

As a result, at the EP3 located at $\gamma_0=\sqrt{2}\Omega_R$ and $k=0$, all three energy bands completely close, as depicted in Fig.\ref{bandfig}(d). This closure of the energy bands is reflected by the behavior where the population of the state $\ket{1}$ remains constant, regardless of the passage of time. This static population indicates the degeneracy at the EP3.

\vspace{10pt}
\paragraph*{\bf PT-symmetry breaking across EP3} Especially, Eq.(\ref{hamiltoniank0eq}) represents the Hamiltonian of system at $k=0$. The eigenspectrum is given by $\lambda+i\frac{\gamma_0}{2}=0, \pm\sqrt{\left(\frac{\sqrt{2}\Omega_R}{2}\right)^2-\left(\frac{\gamma_0}{2}\right)^2}$, which, when adjusted by subtracting a constant complex number, reveals spectral symmetries $\{E_i\}=\{-E_i\}$ and $\{E_i\}=\{{\pm}E_i^*\}$. $\{E_i\}$ reflects the set of the complex eigenvalues of Eq.(\ref{hamiltoniank0eq}). This symmetries indicate the presence of \emph{PT} symmetry in Eq.(\ref{hamiltoniank0eq})~\cite{budich2019symmetry}. By monitoring the evolution of the state $\ket{1}$ population for varying dissipation rates at $k=0$, we observe an abrupt change near the EP3, which signifies the breaking of the \emph{PT} symmetry. It indicates a phase transition point where the system undergoes a qualitative change in its properties and dynamics~\cite{ozturk2021observation,makris2008beam}. 

Furthermore, at $k=0$ the evolution of the system transitions from oscillatory behavior to damping, which implies the closing of the real parts of the energy bands and the opening of the imaginary parts. This spectral behavior suggests the transformation of the system's dynamics from a coherent oscillation regime to a dissipative regime. This evolution reflects the complex nature of the energy bands and their response to the dissipation rate in a non-Hermitian system~\cite{guo2009observation,feng2017non}. 

Finally, by plotting the band structure in the space of dissipation rate, $\gamma=\gamma_0=\frac{1}{2}\gamma_1$, at $k=0$ as shown in Fig.\ref{bandfig}(a), we can observe a clear \emph{PT} symmetry breaking after the EP3, which highlights the phase transition point. 

\paragraph*{\bf Response Sensitivity to Small Perturbation.}
\begin{figure}[htbp]
    \centering
    \includegraphics[width=1.1\linewidth]{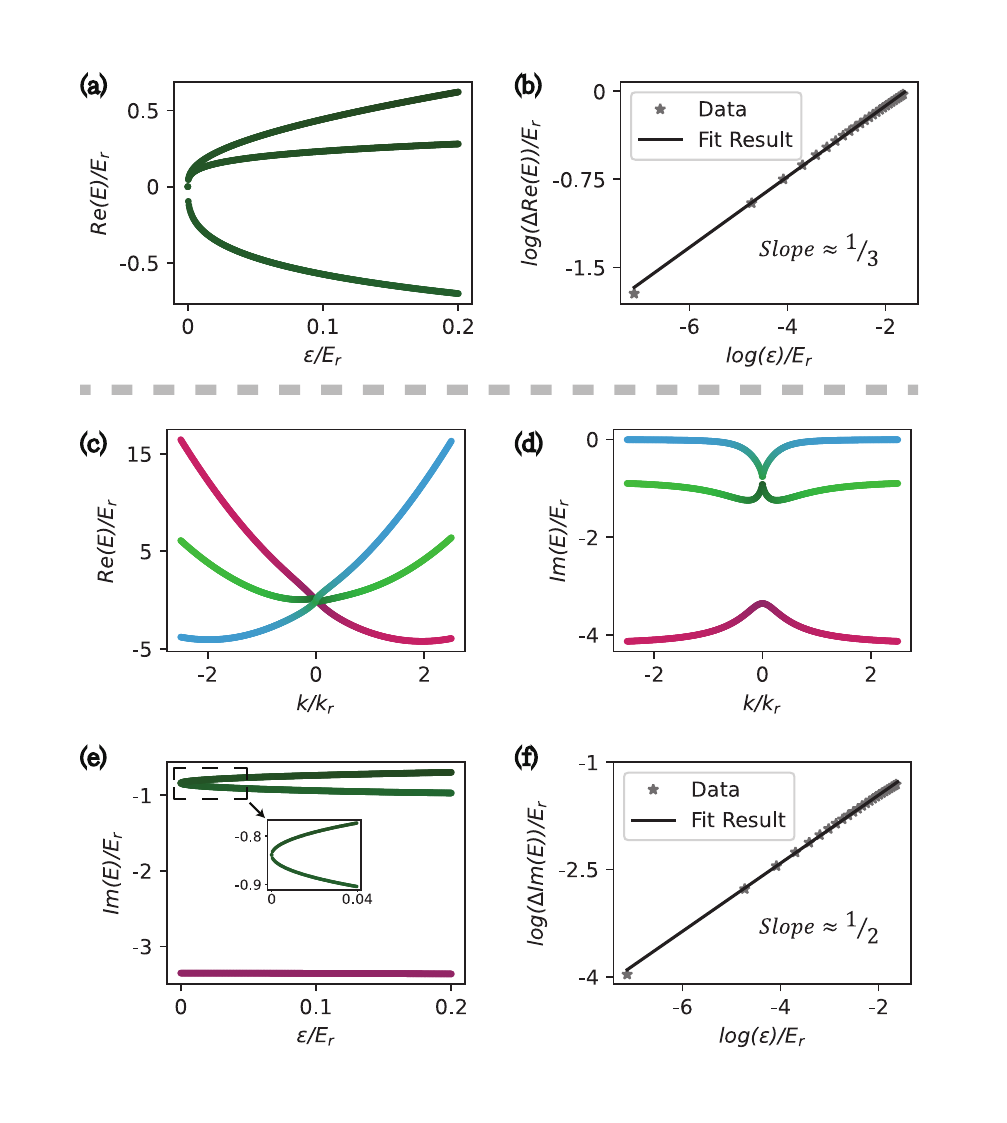}
    \caption{\textbf{Response sensitivity to small perturbation $\epsilon$} (a-b) shows real energy splitting at EP3 when increasing the amplitude of the perturbation $\epsilon$ (left) and the energy splitting between middle and lower energy bands on a logarithmic scale (right). (c-d) Real (left) and imaginary (right) parts of the energy dispersion in quasi-momentum space for $\delta'_1=-\delta'_2=4E_r$, $\Omega'_{R1}=2\Omega'_{R2}=3E_r$ and $\gamma'_1=5\gamma'_0=\sfrac{5\sqrt{2}\Omega'_{R1}}{4}$ at EP2. (e-f) shows the imaginary energy splitting at EP2 when increasing the amplitude of $\epsilon$ (left) and the energy splitting between the collapsed bands on a logarithmic scale (right).}
    \label{sensitivityfig}
\end{figure}

To recognize the order of an EP, we can examine the sensitivity of the response to perturbations in a three-level non-Hermitian system. The sensitivity is typically related to the order of the EP, denoted as $\epsilon^{1/n}$, where $n$ is the order of the EP~\cite{hodaei2017enhanced,wu2021high}.

As mentioned previously, where $\delta_1=-\delta_2=4E_r$, $\Omega_R=\Omega_{R1}=\Omega_{R2}$, and $\gamma_1=2\gamma_0=2\sqrt{2}\Omega_{R1}$ in Hamiltonian (\ref{hamiltonianeq}), an EP3 exists at $k=0$. We can rewrite it as $H=H_0-\sfrac{i\gamma_0}{2}{I}_3$ with the unperturbed Hamiltonian $H_0$ expressed as,
\begin{equation}
    \label{h0eq}
    H_0=\begin{pmatrix}
        -\sfrac{i\gamma_0}{2}&\sfrac{\Omega_R}{2}&0\\
        \sfrac{\Omega_R}{2}&0&\sfrac{\Omega_R}{2}\\
        0&\sfrac{\Omega_R}{2}&\sfrac{i\gamma_0}{2}\\
    \end{pmatrix}.
\end{equation}

In the following, we omit the term $\sfrac{i\gamma_0}{2}$ from the Hamiltonian trace and primarily focus on $H_0$, which will not alter the properties of Hamiltonian (\ref{hamiltonianeq}), except that the eigenvalues are adjusted by subtracting $\sfrac{i\gamma_0}{2}$~\cite{hassan2017dynamically,guo2009observation}.

To assess the response sensitivity of the system at EP3, we introduce a small perturbation $\epsilon$ to the state $\ket{1}$. This small perturbation $\epsilon$ can be experimentally realized by perturbing the parameter $\delta_1$. Consequently, the characteristic equation of $H_0$ with the perturbation $\epsilon$ takes the form:
\begin{equation}
    \label{charep3eq}
    -\lambda\left(\lambda^2+\sfrac{\gamma_0^2}{4}-\sfrac{\Omega_R^2}{2}\right)+\epsilon\left(\lambda^2-\sfrac{i\gamma_0}{2}\lambda-\sfrac{\Omega_R^2}{4}\right)=0.
\end{equation}

From Eq.(\ref{charep3eq}), we can get the expression of perturbation $\epsilon$ at EP3 with $\gamma_0=\sqrt{2}\Omega_R$,
\begin{equation}
    \label{epsilonep3eq}
    \epsilon=\lambda^3\frac{1}{\lambda^2-\sfrac{i\gamma_0}{2}\lambda-\sfrac{\Omega_R^2}{4}}.
\end{equation}

From Eq.(\ref{epsilonep3eq}), we can derive the cubic-root energy dispersion $\lambda\propto\epsilon^\frac{1}{3}$ in the vicinity of EP3 while $\lambda$ approaches the degenerate eigenvalues at EP3~\cite{wu2021high}. 

Due to the non-zero determinants of the principal submatrix $H_{11,22,33}$, the response order of EP3 to the perturbation remains independent of which spins experience the perturbation $\epsilon$~\cite{wu2021high}. Therefore, irrespective of the specific nuclear spin to which the perturbation is applied, we can always observe a response relation $\lambda\propto\epsilon^\frac{1}{3}$ for $\epsilon\ll1$.

From Fig.\ref{sensitivityfig}(a-b), we observe the splitting of the eigen-energy as a function of the small perturbation $\epsilon$. Specifically, we focus on the real part of the energy difference between the middle and lower bands and record it as a function of $\epsilon$. By plotting this data in a logarithmic scale, as depicted in Fig.\ref{sensitivityfig}(b), we can observe a linear relationship. The slope of this linear fit is approximately $\sfrac{1}{3}$, which is consistent with the expected behavior described by Eq.(\ref{epsilonep3eq}) for $\epsilon\ll1$. This finding provides further evidence for the relation $\lambda\propto\epsilon^\frac{1}{3}$ at the EP3, demonstrating the sensitivity of the system's response to small perturbation $\epsilon$~\cite{hodaei2017enhanced}.

It would be interesting to compare the two-fold degenerate case with EP3. In this case, we modify the parameters of the Hamiltonian from the previous values to $\Omega'_{R}=\Omega'_{R1}=2\Omega'_{R2}$ and $\gamma'_1=5\gamma'_0$. At $k=0$, we set $\delta'_1=-\delta'_2=4E_r$. The Hamiltonian is transformed into the following form~\cite{laha2020third},
\begin{equation}
    \label{ep2eq}
    H^{2fold}_0-\sfrac{i\gamma'_0}{2}I_3=\begin{pmatrix}
        -i2\gamma'_0&\sfrac{\Omega'_R}{2}&0\\
        \sfrac{\Omega'_R}{2}&0&\sfrac{\Omega'_R}{4}\\
        0&\sfrac{\Omega'_R}{4}&\sfrac{i\gamma'_0}{2}\\
    \end{pmatrix}-\sfrac{i\gamma'_0}{2}I_3.
\end{equation}

Using the same color scale for the spin population as shown in Fig.\ref{bandfig}, we visualize the band structure of the Hamiltonian (\ref{ep2eq}) in Fig.\ref{sensitivityfig}(c-d). At $k=0$, two of the bands collapse, resulting in an eigenvalue of $-\sfrac{i\gamma_0}{2}$ and an eigenstate of $\left(\sfrac{\sqrt{10}}{10}, \sfrac{i\sqrt{2}}{2}, -\sfrac{\sqrt{10}}{5}\right)^T$, which is similar to the EP2 in a two-band system. This collapse of the bands at EP at $k=0$ can be clearly observed in the band structure plot Fig.\ref{sensitivityfig}(c-d). The energy spectrum of Eq.(\ref{ep2eq}) is given by $\lambda-i\frac{\gamma'_0}{2}=0,-i\frac{3\gamma'_0}{4}\pm\sqrt{\left(\frac{\sqrt{5}\Omega_R}{4}\right)^2-\left(\frac{5\gamma'_0}{4}\right)^2}$. When adjusted by subtracting the constant complex part $i\frac{\gamma'_0}{2}$, it retains only the anti-conjugate spectral symmetry $\{E_i\}=\{-E_i^*\}$.

For a three-level quantum system, achieving EP3 in 2D parameter space demands specific symmetry conditions, particularly \emph{PT}-symmetry in our system. According to the discriminant of Cardano formula, before reaching EP3, there exist EP2s which form so-called exceptional nexus in the parameter space~\cite{tang2020exceptional}. In this case, we only consider $H^{2fold}_0$, with the trace subtracted. Similar to the EP3 scenario, the non-zero submatrix of $H^{2fold}_0$ ensures that the system exhibits consistent response sensitivity regardless of the specific nuclear spin being perturbed. Thus, we apply a small perturbation with $\epsilon\ll1$ to the state $\ket{1}$. Consequently, the characteristic equation of $H_0^{2fold}$, as described in Eq.(\ref{ep2eq}), can be expressed as follows,

\begin{equation}
    \label{characep2eq}
    \begin{aligned}
        -\lambda\left(\lambda^2+\sfrac{i3\gamma'_0}{2}\lambda+{\gamma'}_0^2-\left(\sfrac{\sqrt{5}\Omega'_R}{4}\right)^2\right)&\\+\epsilon\left(\lambda^2-\sfrac{i\gamma'_0}{2}\lambda-\left(\sfrac{\Omega'_R}{4}\right)^2\right)&=0.
    \end{aligned}
\end{equation}

The expression for the perturbation $\epsilon$ at EP with $\gamma'_0=\sfrac{\sqrt{5}\Omega'_R}{4}$ can be written as, 
\begin{equation}
    \label{epsilonep2eq}
    \epsilon=\lambda^2\frac{\lambda+\sfrac{i3\gamma'_0}{2}}{\lambda^2-\sfrac{i\gamma'_0}{2}\lambda-\left(\sfrac{\Omega'_R}{4}\right)^2}.
\end{equation}

Considering the energy bands closely connecting to the EP2, it is observed that there exists a square-root energy dispersion $\lambda\propto\epsilon^\frac{1}{2}$ as $\lambda$ approaches the eigenvalue of EP~\cite{tang2020exceptional,wu2021high}.

However, as shown in Fig.\ref{sensitivityfig}(c-d), the real parts of the eigenvalues all collapse to zero at $k=0$ for EP while the imaginary parts of the eigenvalues of EP and the remaining one are significantly separated. 

Therefore, we extract the splitting of the imaginary part of the energy difference between the two collapsed bands where EP exists, as depicted in Fig.\ref{sensitivityfig}(e-f). It shows the relation between the splitting and the varying perturbation $\epsilon$ is consistent with the square-root energy dispersion $\lambda\propto\epsilon^\frac{1}{2}$ for $\epsilon\ll1$, the conclusion of which holds true for the real part as well.

By comparing the results obtained from EP3 with cubic-root energy dispersion and EP with square-root energy dispersion, we can conclude that the sensitivity of a system at EPn in response to the perturbation $\epsilon$ is not only proportional to $\epsilon^\frac{1}{N}$~\cite{tang2020exceptional,wu2021high} but also closely related to the degeneracy and the symmetry of the system. Since we consider the case where $\epsilon\ll1$, higher-order EPs result in greater sensitivity of the system to small perturbation $\epsilon$ at EP. This characteristic behavior of EP highlights the distinct nature of EPs with different symmetries and order.

\begin{figure}[htbp]
    \centering
    \includegraphics[width=1\linewidth]{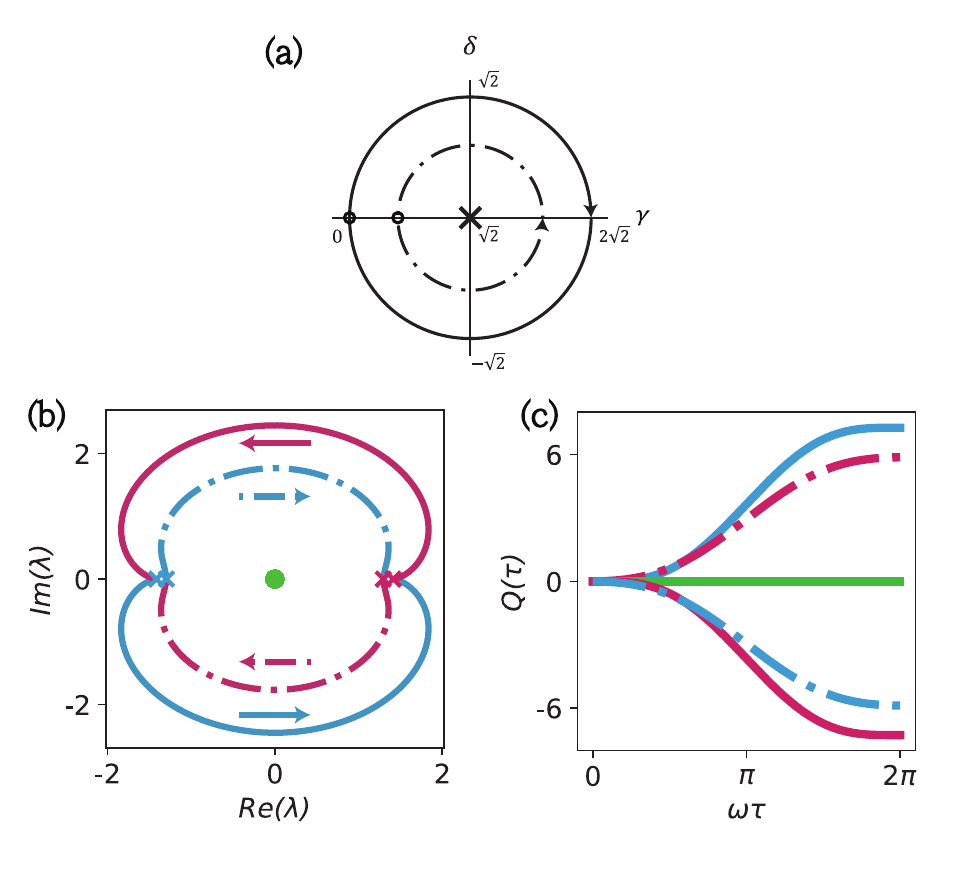}
    \caption{\textbf{Dynamics of quasi-static encircling process} Encircling path in the parameter space is shown in (a). (b) represents the eigenvalue trajectories when quasistatically encircling EP3 along the parametric loop shown in (a) and the eigenvalues $\lambda_{+1,0,-1}$ are distinguished by red, blue and green colors, respectively. The starting points are marked with coloured $\times$. Path directions are shown with arrows and linestyle. The solid line and dash-dotted line correspond to the CW and CCW encircling directions, respectively. (c) plots the accumulated phase $Q$ during the encirclement associated with different eigenvalue trajectories.}
    \label{quasifig}
\end{figure}

\begin{figure}
    \centering
    \includegraphics[width=1\linewidth]{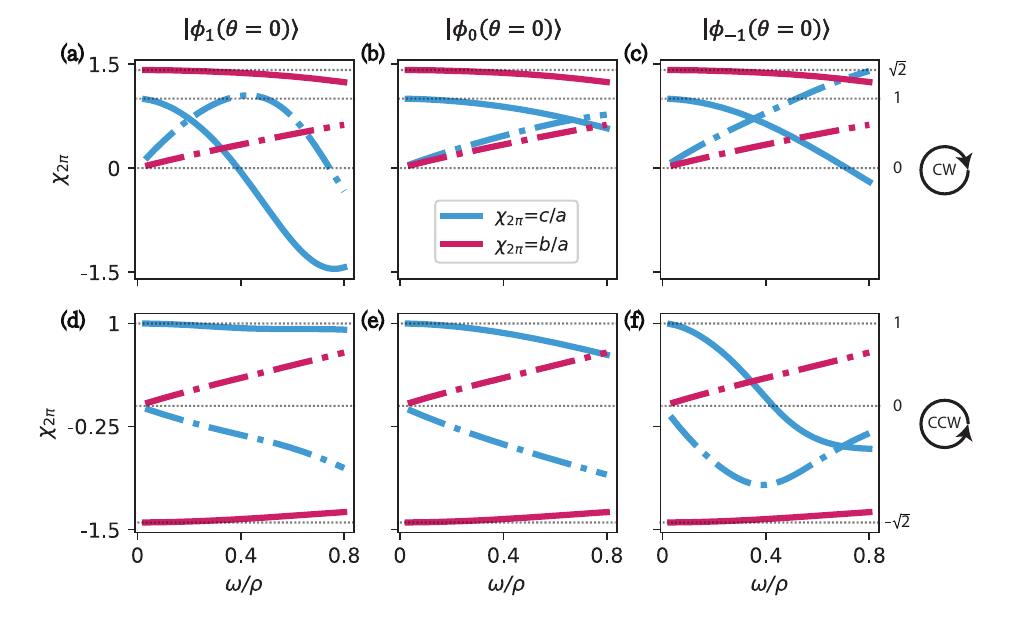}
    \caption{\textbf{Final states after single encircling of EP3 for different degree of adiabaticity} The upper (a-c) and lower (d-f) panels correspond to CW and CCW encircling. From left to right, the inputs are chosen as $\ket{\phi_1}$, $\ket{\phi_0}$ and $\ket{\phi_{-1}}$, respectively. The solid line and dash-dotted line are associated with the real and imaginary parts of the ratios $\chi_{2\pi}=c/a$ (blue) and $\chi_{2\pi}=b/a$ (red) of final states.}
    \label{encirclefig}
\end{figure}

\vspace{20pt}
{\bf Encircling dynamics.} Here we consider the Hamiltonian $H_0$ with the trace subtracted in the parametric space at $k=0$. To explore the encircling of EP3 in the parametric space, we fix certain parameters to their respective values in the Hamiltonian (Eq.(\ref{h0eq})), while allowing the dissipation rates and two-photon detuning, $\delta_1$ and $\delta_2$, to vary with time. Specifically, we impose the conditions $\gamma_0=\sfrac{\gamma_1}{2}$ and $\delta_1=\delta_2+8E_r$ during the encirclement process to simplify the encircling procedure and provide a more manageable framework for studying the EP3 in the parametric space.

In this arrangement, the dynamics of the system are characterized in the Schr\"odinger description by $i\partial_t\ket{\phi(t)}+H_0(t)\ket{\phi(t)}=0$, which holds true for both adiabatic and nonadiabatic cases~\cite{hassan2017dynamically}. Here, $\ket{\phi(t)}=\left(a(t), b(t), c(t)\right)^T$ represents the time-dependent state vector, and $H_0(t)$ is a concise form of the time-dependent Hamiltonian that captures the relevant dynamics of the system,
\begin{equation}
    \label{timehamiltonianeq}
    H(t)=
    \begin{pmatrix}
        -\tilde{\delta}(t)-i\tilde{\gamma}(t)&\sfrac{\Omega_R}{2}&0\\
        \sfrac{\Omega_R}{2}&0&\sfrac{\Omega_R}{2}\\
        0&\sfrac{\Omega_R}{2}&\tilde{\delta}(t)+i\tilde{\gamma}(t)\\
    \end{pmatrix}
\end{equation}
where $\tilde{\delta}=-4E_r+\delta_1=4E_r+\delta_2$ and $\tilde{\gamma}=\sfrac{\gamma_0}{2}$. 

For convenience, we henceforth scale the variables $(\sfrac{2\tilde{\delta}}{\Omega_R}, \sfrac{2\tilde{\gamma}}{\Omega_R}, \sfrac{\Omega_Rt}{2})\to(\delta, \gamma, \tau)$ for the sake of simplicity. In the parametric space defined by $\delta$ and $\gamma$, the EP3 is located at $\delta=0$ and $\gamma=\sqrt{2}$. By following a circular trajectory in the parametric space, we can effectively encircle the EP3,
\begin{equation}
    \label{circleeq}
    \begin{cases}
        \gamma=\sqrt{2}\left(1-{\rho}cos(\omega{\tau})\right)\\
        \delta=\sqrt{2}{\rho}sin(\omega{\tau})\\
    \end{cases}
\end{equation}
where $\rho$ denotes the radius of the encircling circle in the $\delta-\gamma$ parametric plane ($\rho\ll1$) and $\omega$ represents the speed of the encircling process, which reflects the adiabaticity of the process. Eq.(\ref{circleeq}) indicates a clockwise (CW) parametric circle if $\omega>0$ and a counterclockwise one if $\omega<0$~\cite{hassan2017dynamically}.

In order to gain a comprehensive understanding of the state evolution along the parametric circle described by Eq.(\ref{circleeq}), we can break down Eq.(\ref{timehamiltonianeq}) into a system of fourth-order differential equations for the state variables $a(\tau)$ and $c(\tau)$ of $\ket{\phi(\tau)}$ as follow,
\begin{subequations}
    \label{differentialeq}
    \begin{equation}
        \label{differentialaeq}
        \partial_\tau\left(\partial_\tau^3a-2\Sigma_a\partial_{\tau}a-\partial_\tau\Sigma_aa\right)=0
     \end{equation}
    \begin{equation}
        \label{defferentialceq}
        \partial_\tau\left(\partial_\tau^3c+2\Sigma_c\partial_{\tau}c+\partial_\tau\Sigma_cc\right)=0
    \end{equation}
\end{subequations}
where $\Sigma_{a, c}=\sqrt{2}\partial_\tau\Delta\pm\Delta^2\mp1$ and $\Delta=1-\rho{e}^{i\omega\tau}$.

To encircle the EP3 once, the trajectory starts at $\tau=0$ and ends at $\tau=2\pi$. By setting $sin\theta=(1-\rho)$, the eigenvalues at the endpoints of the trajectory are given by $\lambda_0=0$ and $\lambda_{\pm1}=\pm\sqrt{2}cos\theta$ which correspond to the eigenvectors $\ket{\phi_{\pm1}}\propto(\sfrac{e^{\mp{i\theta}}}{2}, \sfrac{\pm\sqrt{2}}{2}, \sfrac{e^{\pm{i\theta}}}{2})^T$ and $\ket{\phi_0}\propto(1, i\sqrt{2}sin\theta, -1)^T$. 

To gain insights into the encircling behavior under the stationary condition, we can begin by observing the quasistatic motion of the system during the encirclement around the EP3~\cite{hassan2017dynamically}. Fig.\ref{quasifig}(a-b) illustrates that the eigenvalues except the zero mode undergo a swap with each other at the end of the encirclement. 

Similar to the EP2 case, the encircling direction determines the dominance of one of the two non-zero modes, reflected in the amplifications $e^{Q(\tau)}$ of each mode, as shown in Fig.\ref{quasifig}(c), where $Q(\tau)=-\int_0^\tau{dt'}Im[\lambda(t')]$. The quantity $Q(\tau)$ represents the accumulated phase due to the imaginary part of the eigenvalue $\lambda(t')$ over the course of the encirclement. By examining the amplifications of the modes, we can analyze the importance of each mode during the encircling process~\cite{hassan2017dynamically}.

However, a stationary analysis is insufficient to capture the dynamical evolution which follows the Eqs.(\ref{differentialeq}). To gain a more precise understanding, we numerically solve the Eqs.(\ref{differentialeq}) for different encircling directions (CW/CCW) and inputs ($\ket{\phi_{0, \pm1}}$). The resulting numerical solutions are shown in Fig.\ref{encirclefig}, where the ratios of the state variables at the terminal of parametric circle are exhibited. These numerical solutions allow us to observe the behavior of the state variables as they approach the terminal points of the parametric circle.

At $\rho=1$, the inputs in Fig.\ref{encirclefig} correspond to $\ket{\phi_0}=\left(\sfrac{\sqrt{2}}{2}, 0, -\sfrac{\sqrt{2}}{2}\right)^T$ and $\ket{\phi_{\pm1}}=\left(\sfrac{1}{2}, \pm\sfrac{\sqrt{2}}{2}, \sfrac{1}{2}\right)^T$. By scrutinizing the near-adiabatic results for small $\omega$, we observe that if the encirclement around EP3 is performed once in a CW manner, as shown in Fig.\ref{encirclefig}(a-c), the system will end in $\ket{\phi_{+1}}$ multiplied by a global phase of no physical consequence, irrespective of the inputs according to the ratios $\chi_{2\pi}$ of the final state. Conversely, for CCW encirclement, as demonstrated in Fig.\ref{encirclefig}(d-f), the system will end in $\ket{\phi_{-1}}$.

Although the non-zero modes of EP3 exhibit similar quasistatic manner to those of EP2, they differ in their dynamical processes. Fig.(\ref{encirclefig}) suggests that in the adiabatic limit, the outcome of a single encircling around EP3 is direction-dependent and dominated by a specific non-zero mode, regardless of the inputs, which is similar to the case of EP2~\cite{doppler2016dynamically,hassan2017chiral}.

However, as the encircling speed $\omega$ increases, the system's evolution becomes less likely to follow the trajectory of the adiabatic case. This transition from the adiabatic limit to the nonadiabatic case leads to the elimination of the aforementioned mode conversion phenomenon. This behavior is similar to what is observed in the case of EP2. However, it is worth noting that the adiabatic limit of EP2 is significantly higher than that of EP3, which implies that the mode conversion for the encirclement of EP3 only occurs when the encircling speed $\omega$ is sufficiently small.

Moreover, the nonadiabatic evolution of the encirclement of EP3 is highly dependent on the initial conditions, which can be witnessed by the relationship between $\chi_{2\pi}=c/a$ and $\omega$ with $\omega$ larger than the adiabatic limit, as demonstrated in Fig.\ref{encirclefig}. The final state obtained after nonadiabatically encircling EP3 once varies depending on the different initial states $\ket{\phi_{0, \pm1}}$. This is distinct from the EP2, where the initial state in the expression of the ratio $\chi_{2\pi}$ cancels out, resulting in a consistent final state regardless of the initial conditions~\cite{hassan2017dynamically,alma991000442119703412}. Thus, the nonadiabatic behavior of the encircling of EP3 is characterized by a sensitivity to the initial conditions, leading to different final states depending on the chosen initial states $\ket{\phi_{0, \pm1}}$.

{\bf Conclusions}
We theoretically investigate the EP3 in a dissipative three-level spin-orbit-coupled fermions system. Our study covers various aspects, including the energy band structure in momentum space, the system's response sensitivity to small perturbation $\epsilon$, and the encirclement behaviors. We compared the characteristics of EP3 with those of EP2. Moreover, we also propose an experimental scheme utilizing ultracold $^{173}$Yb fermions to explore. 

By carefully adjusting the the dissipation rates and the two-photon detuning, it is possible to engineer a scenario where all three energy bands of the system collapse at one point, which corresponds to the three-fold degeneracy in both eigenvalues and eigenstates at $k=0$. 

Additionally, we introduce a small perturbation $\epsilon$ by varying the two-photon detuning $\delta_1$ to observe the splitting at EP3. As a result, the energy splitting at EP3 with respect to the small perturbation $\epsilon$ is found to be proportional to $\epsilon^{1/3}$ when examining it on a logarithmic scale. For comparison, for EP2, achieved by adjusting the coupling strengths and dissipation in the three-level system, the energy splitting is proportional to $\epsilon^{1/2}$. The different scaling exponents reflect the distinct degeneracy structures and underlying symmetry of EP3 and EP2, respectively. These findings provide opportunities for controlling and manipulating systems near these degeneracy points using external perturbations.

Regarding the encirlcing around the EP, EP3 exhibits similar encircling properties for its non-zero modes in both quasistatic and adiabatic situations, similar to EP2. However, in the nonadiabatic case, EP3 and EP2 differ in terms of the final state after a single encirclement. In the case of EP3, the final state not only depends on the encircling direction but also on the inputs. On the other hand, for EP2, the final state is solely determined by the encircling direction, regardless of the inputs. Our work hints at intriguing characteristics of higher-order exceptional points, which could potentially be realized in optical lattices~\cite{song2018,Song2019,Zhao.2022,Zhao.2023}.

\vspace{10pt}
\paragraph*{\bf Acknowledgement}
GBJ acknowledges support from the RGC through  16302821, 16306321, 16306922, 16302123, 16305024, C6009-20G, N-HKUST636-22, and RFS2122-6S04. CH acknowledges support from the RGC for RGC Postdoctoral fellowship.

\bibliography{EP3_v1.bib}% Produces the bibliography via BibTeX.
\end{document}